\documentclass[10pt,a4paper,twocolumn,superscriptaddress,aps,nofootinbib ]{revtex4}
\usepackage{amsmath,amssymb,graphicx}
\makeatletter
\usepackage[active]{srcltx}
\usepackage{graphicx,color}
\usepackage{changebar}
\usepackage{hyperref}
\hypersetup{
	colorlinks=true,
	urlcolor= blue,
	citecolor=blue,
	linkcolor= blue,
	bookmarks=true,
	bookmarksopen=false,
}
\usepackage[T1]{fontenc}
\usepackage{ae}
\usepackage[ansinew]{inputenc}
\usepackage{amsmath}
\usepackage{braket}
\usepackage{mathtools}
\usepackage{slashed}
\usepackage{empheq}
\usepackage{tikz}
\usepackage{multirow}
\usetikzlibrary{decorations.pathmorphing}
\usepackage{amsmath,amssymb,graphics,epsfig,subfigure}
 \usetikzlibrary{quotes,angles}
 \usepackage{etoolbox}
 
\usepackage{graphicx}
\usepackage{amsfonts}
\usepackage{amsmath}
\usetikzlibrary{arrows} 
\usetikzlibrary{decorations.markings}

\newcommand{\orcidicon}{%
	\begin{tikzpicture}
	\draw[lime, fill=lime] (0,0) 
	circle [radius=0.15] 
	node[white] {{\fontfamily{qag}\selectfont \tiny ID}};
	\draw[white, fill=white] (-0.0625,0.095) 
	circle [radius=0.007];
	\end{tikzpicture}	\hspace{-2mm}
}
\newcommand\orcidAdriano{{\href{https://orcid.org/0000-0003-1871-2068}{\orcidicon}}}
\newcommand\orcidCarlos{{\href{https://orcid.org/0000-0001-6913-0223}{\orcidicon}}}
\newcommand\orcidRicardo{{\href{https://orcid.org/0000-0001-8802-3634}{\orcidicon}}}

\begin{document}

\title{ Gravitational lensing in a topologically charged  Eddington-inspired Born-Infeld spacetime}

\author{A. R. Soares\orcidAdriano\!}
\email{adriano.soares@ifma.edu.br}
\affiliation{Grupo de Estudos e Pesquisas em Laborat\'orio de Educaç\~ao matem\'atica, Instituto Federal de Educa\c{c}\~ao Ci\^encia e Tecnologia do Maranh\~ao,  R. Dep. Gast\~ao Vieira, 1000, CEP 65393-000 Buriticupu, MA, Brazil.}

\author{R. L. L. Vit\'oria \orcidRicardo\!}
\email{ricardo.vitoria@pq.cnpq.br/ricardo-luis91@hotmail.com}
\affiliation{Faculdade de F\'isica, Universidade Federal do Par\'a, Av. Augusto Corr\^ea, Guam\'a, 66075-110, Bel\'em, PA, Brazil.}

\author{C. F. S. Pereira \orcidCarlos\!}
\email{carlos.f.pereira@edu.ufes.br}
\affiliation{Departamento de F\'isica e Qu\'imica, Universidade Federal do Esp\'irito Santo, Av.Fernando Ferrari, 514, Goiabeiras, Vit\'oria, ES 29060-900, Brazil}

\begin{abstract}
	
In the present paper, we study several aspects of gravitational lensing caused by a topologically charged Monopole/Wormhole, both in the weak field limit and in the strong field limit. We calculate the light deflection and then use it to determine the observables, with which one can investigate the existence of these objects through observational tools. We emphasize that the presence of the topological charge produces changes in the observables in relation to the case of General Relativity Ellis-Bronnikov wormhole.
	
\end{abstract}



\maketitle

\section{Introduction}

It is known that despite its enormous success,  the Einstein's General Relativity (GR) has some weaknesses that may point to a broader gravitational theory. In particular, we highlight the singularity problem, i.e., the termination of a geodesic in a black hole singularity. The theory also does not naturally deal with the accelerating expansion of the universe and the cosmological problem of the Big Bang singularity \cite{I1,I2,I3,I4,I5,I6}. These, among other questions, motivated physicists to search for a theory capable of avoiding such problems \cite{f1,f2,f3,f4,f5}. Among these theories, we highlight the so-called  Eddington-inspired Born-Infeld modification of gravity (EiBI gravity)\cite{f5}. The structure of this theory is inspired by the nonlinear electrodynamics of Born and Infeld \cite{E1} and, when approached by the affine-metric formalism, avoids problems such as phantom degrees of instability \cite{olmo2011}. Even at a classical level (without quantum corrections), such a theory provides  singularity-free black holes solutions and sustained wormholes without the need for exotic matter \cite{o1,o2,o3,Soares2020, Soares2019}. In fact, these new theoretical possibilities, combined with the technological tools arising from large international collaborations, have rekindled the search for regular black holes and wormholes.

One of the forms of investigation in cosmology and gravitation are gravitational lensing \cite{Liebes,Mellier1999,Schneider2001,Kaiser1993,Chowdhuri174}, Einstein himself used this method to make GR effectively relevant in the scientific community \cite{Einstein1936,Crispino2019}. Gravitational lensing can occur both in the weak field limit, which is when the light ray passes very far from the source responsible for the lens, or in the so-called strong field limit, which is when the light passes very close, in this case the deflection of light it is actually very big \cite{Atkinson1965,Darwin}. Although the mathematical treatment of lensing in the strong field regime is more complicated, recent works have made it possible to treat the subject even analytically. In \cite{Boz-Cap2001,Bozza2002}, the authors obtained an expression to calculate the deflection of light in the strong field limit and applied it to the case of the Scharszchild spacetime. In \cite{Tsukamoto2017-2}, Tsukamoto presents an improved expression of light deflection in static and asymptotically flat spacetimes as well in ultrastatic spacetimes. In \cite{Virbhadra-Ellis-2000},  Virbhadra and Ellis showed that in the strong field limit there is, in addition to the first and second images, an infinity of images called relativistic images that do not discriminate between themselves. In \cite{Perlick2004}, Perlick considered gravitational lensing exactly in a spherically symmetric and static spacetime, exemplifying with  o Barriola-Vilenkin monopole and  Ellis-Bronnikov wormhole. Since then, gravitational lensing has been investigated in several contexts involving black holes \cite{Eiroa2002, Eiroa2004, Tsukamoto2017, rotacao, Aazami, azquez, Bozza-Sereno, Bozza-Scarpetta, VBz,Virbhadra-064038, Virbhadra-2204.01792}, wormholes \cite{Nakajima85, Chetouani, Nandi, Dey-sen, Gibbons-Vyska, Tsukamoto-Harada, Nandi-Potapov, Tsukamoto2017-95, Tsukamoto-Harada-2017, Shaikh-Banerjee, Gao2211.17065, Huang2302.13704,Abe2010}, topological defects \cite{Cheng, Sharif2015, Cheng92, Cheng28}, modified theories of gravitation \cite{Sotani92, Wei75, Bhadra, Eiroa73, Sarkar23, Mukherjee39, Gyulchev75, Chen80, Shaikh96} and regular black holes \cite{Eiroa28,Eiroa88,soaresBbounce, Ghosh006}.

So far, the simplest solution ever obtained in EiBI gravity is the Global Monopole/Wormhole (GM/WH for short) \cite{Soares2020,lambaga2018}. This solution interpolates a modified global monopole  and a wormhole like Ellis-Bronnikov with topological charge. We must emphasize that said solution was obtained with a source of matter that does not violate the energy conditions and, impressively, it is as simple as the well-known Ellis-Bronnikov solution \cite{ellis,bronn}. Given its simplicity and potential applications to understand systems linked to condensed matter, some studies related to the quantum dynamics of particles in this space-time have already been published \cite{me2, me3, CARB, Ahmed141, Aounallah1013}. With regard to gravitational lensing, in \cite{lambaga2018}, the authors calculated the deflection of light in the EiBI-GM in the weak field limit and in \cite{Soares2020} the authors obtained a general expression for the light deflection in the topologically charged WH. 

In the present paper, we will do a more general study of the topologically charged EiBI spacetime. Let's consider both possibilities, WH and GM, and let's treat them in the same theoretical framework. We will address the deflection of light not only in the weak field limit, but also in the strong field limit for the first time. Furthermore, we calculate the observables and discuss the possibility of their detection. It should also be noted that the results obtained here can also shed light on the optical properties of liquid crystals and crystalline lattices with topological defects \cite{Moraes103}. 

 The paper is organized as follows: In section \ref{sec2} we present the metric that describes the GM/WH in EiBI gravity and calculate the light deflection in the weak field limit. In section \ref{sec3} we calculate the light deflection in the strong field limit and investigate the lens equation only in the wormhole spacetime.  In section \ref{sec4}, we present some observational perspectives of the wormhole and its detection plausibility and, finally, in section \ref{conc}, we make the last discussions and conclusions of the paper, presenting our future perspectives.

\section{The GM/WH metric and Lensing in the Weak Field Limit}\label{sec2}
The line element describing the GM/WH in EiBI gravity,
in spherical coordinates ($t,r,\theta$, $\phi$), is given by \cite{Soares2020}: 
\begin{eqnarray} \label{eq1}
	ds^2&=&-(1-\kappa^2\eta^2)dt^2+\frac{r^2dr^2}{(1-\kappa^2\eta^2)(r^2+\varepsilon\kappa^2\eta^2)}\nonumber\\
	&& + r^2(d\theta^2+\sin^2\theta d\phi^2) \ ,
\end{eqnarray}
where $\kappa^2=8\pi G$, being $G$ the gravitational constant, $\eta$ the energy scale associated with spontaneous symmetry breaking and $\varepsilon$, the parameter associated with the non-linearity of the EiBI gravity. For simplicity, let us make just a few reparametrizations: $t\to t\sqrt{1-\kappa^2\eta^2}$, $\alpha=\sqrt{1-\kappa^2\eta^2}$  and $\varepsilon\to \varepsilon(1-\alpha^2)$. With that, the Eq.(\ref{eq1}) becomes
\begin{equation}\label{eq2}
	ds^2=-dt^2+\frac{dr^2}{\alpha^2\big(1+\frac{\varepsilon}{r^2}\big)}+ r^2(d\theta^2+\sin^2\theta d\phi^2) \ .
\end{equation}
 The above metric can describe a global monopole, when $\varepsilon>0$, or a topologically charged wormhole, when $\varepsilon<0$. It should be noted that solution (\ref{eq2}) was originally obtained, in \cite{Soares2020,lambaga2018}, from a energy source corresponding to the  external region of the  Barriola-Vilenkin global monopole core  \cite{Barriola-Vilenkin}. For that reason, in the case $\varepsilon>0$, (\ref{eq2}) also describes a spacetime external to the global monopole core, but this time within the EiBI gravity. And, when $\varepsilon=0$, the metric reduces to on the well-known outside core global monopole. For  $\varepsilon<0$, we can compare the Eq.(\ref{eq2}) with the more general case of a wormhole, proposed by Morris and Thorne \cite{MorrisThorne}, to conclude that the redshift function is null, $\Phi(r)=0$,
  and the form function is $b(r)=r-\alpha^2\big(r-\frac{|\varepsilon|}{r}\big)$. Because of topological charge, the EiBI wormhole is not asymptotically flat, unlike the Ellis-Bronnikov wormhole: $\lim\limits_{r\to\infty}\frac{b(r)}{r}=1-\alpha^2$.
 
 Let us now get the geodesics associated with spacetime (\ref{eq2}), using the variational method. For a smooth curve on a space with metric (\ref{eq2}), the length, $S$, of the curve is
 \begin{equation}\label{eq3}
 	S=\int \sqrt{\bigg(g_{\mu\nu}\frac{dx^\mu}{d\lambda}\frac{dx^\nu}{d\lambda}\bigg)}d\lambda \ ,
 \end{equation}
 where $\lambda$ is the affine parameter of the curve. Taking $S$ as an affine parameter  in  (\ref{eq3}), we can show that the curves that minimize (\ref{eq3}), $\delta S=0$, also minimize:
 \begin{equation}
 		\int \bigg(g_{\mu\nu}\frac{dx^\mu}{d\lambda}\frac{dx^\nu}{d\lambda}\bigg)d\lambda \ =\int \mathcal{L} d\lambda \ .
 \end{equation}
 Therefore, for $\theta = \frac{\pi}{2}$, the lagrangian $\mathcal{L}$ becomes:
 \begin{equation}\label{lag}
 	\mathcal{L}=-\bigg(\frac{dt}{d\lambda}\bigg)^2+ \frac{1}{\alpha^2\big(1+\frac{\varepsilon}{r^2}\big)} \bigg(\frac{dr}{d\lambda}\bigg)^2+r^2\bigg(\frac{d\phi}{d\lambda}\bigg)^2 \ .
 \end{equation}
The corresponding Euler-Lagrange equation for the coordinates $t$ and $\phi$ leads to the following conserved quantities
 \begin{equation}\label{qc}
 	E=\frac{dt}{d\lambda}\hspace{1cm} \text{and} \hspace{1cm} L=r^2\frac{d\phi}{d\lambda} \ .
 \end{equation}
In terms of the quantities $E$ and $L$, the lagrangian, Eq.(\ref{lag}), becomes:
 \begin{equation}\label{l2}
 	\mathcal{L}=-E^2+\frac{1}{\alpha^2\big(1+\frac{\varepsilon}{r^2}\big)}\bigg(\frac{dr}{d\lambda}\bigg)^2+\frac{L^2}{r^2} \ .
 \end{equation}
  It is known that for null geodesics, $\mathcal{L}=0$. In this case, the  (\ref{l2}) leads to
  \begin{equation}\label{pt}
  \bigg(\frac{dr}{d\lambda}\bigg)^2=\alpha^2\bigg(1+\frac{\varepsilon}{r^2}\bigg)  \bigg(E^2-\frac{L^2}{r^2}\bigg) \ .
  \end{equation}
  
\begin{figure}[h]
  \centering
  	\includegraphics[width=\columnwidth]{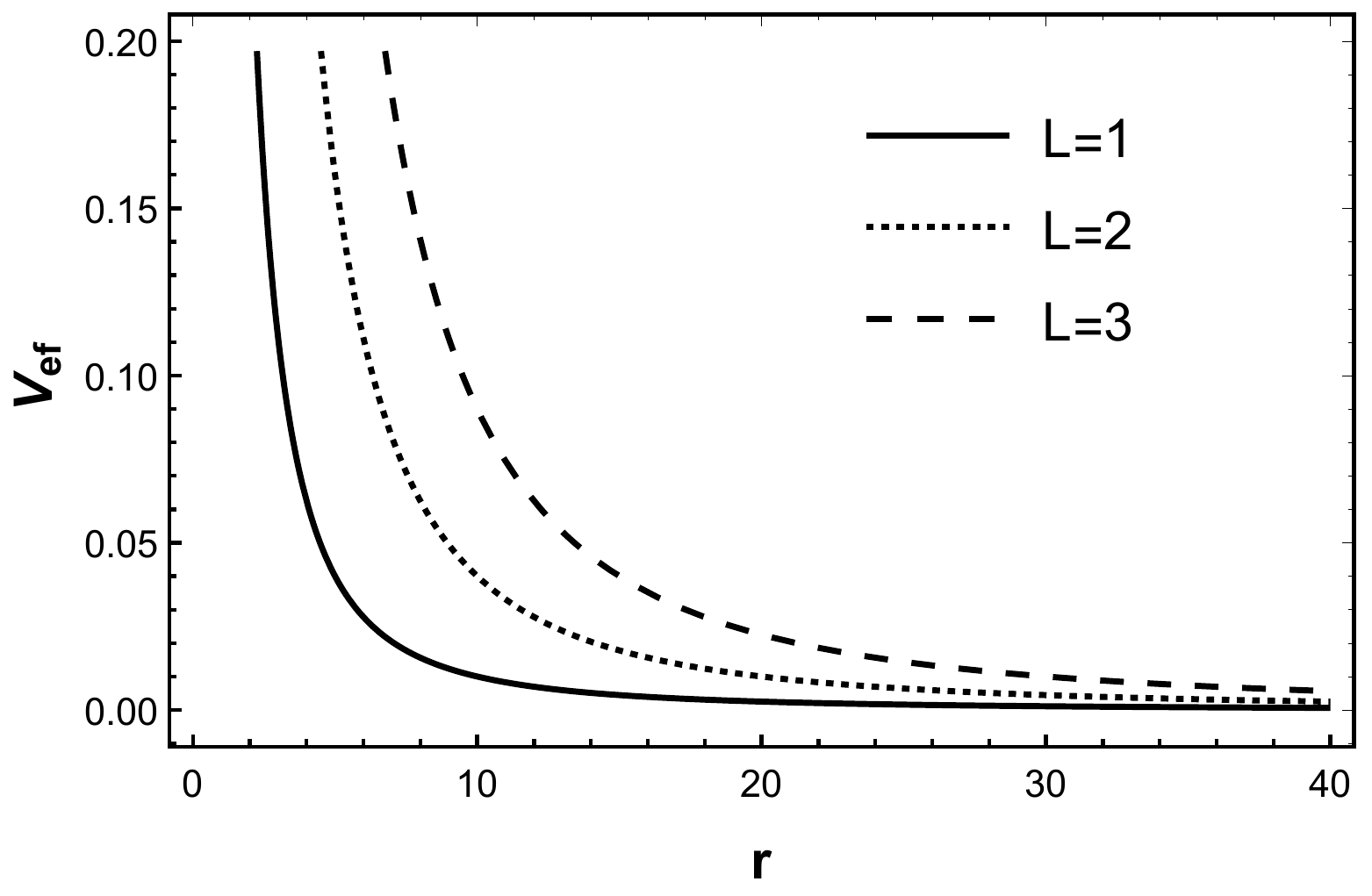}
  \caption{Effective potential for some values of angular momentum $L$.} 
  	\label{I}
  \end{figure}

  This equation can be interpreted as describing the one-dimensional motion of a particle of energy $E$ subject to an effective potential $V_{ef}=L^2/r^2$. In Fig. \ref{I}, we plot the effective potential for some values of $L$. Of course, there is only radial motion when $\frac{dr}{d\lambda}>0$, and the smaller the angular momentum $L$ greater the approximation. Therefore, the turning point $r_0$ occurs for $\frac{dr}{d\lambda}=0$, i.e., $r_0=L^2/E^2$.  It is worth mentioning that in the case of the wormhole, $\varepsilon<0$, the solution has a minimum radius given by $r=\sqrt{|\varepsilon|}$, thus, photons with sufficient energy pass to the other side of the wormhole.
  
  From (\ref{qc}), it can be shown that Eq.(\ref{pt}) becomes
  \begin{equation}\label{2}
  \frac{d\phi}{dr}=\frac{\beta}{\alpha\sqrt{(r^2+\varepsilon)(r^2-\beta^2)}} \ ,
  \end{equation}
  where $\beta=\frac{L}{E}$. We want to find the change in coordinate $\phi$, i.e., $\Delta\phi= \phi_{-}-\phi_{+}$.  By symmetry, the contributions to $\Delta\phi$ before and after the turning point are equal, so
   Eq.(\ref{2}) leads to
  \begin{equation}\label{del}
  	\Delta\phi=2\int_{\beta}^{\infty} \frac{\beta \ dr}{\alpha\sqrt{(r^2+\varepsilon)(r^2-\beta^2)}} \ .
  \end{equation}
 Let us define the variables
  \begin{equation}\label{red}
  t=\frac{\beta}{r} \hspace{1cm}  \text{and} \hspace{1cm} m=-\frac{\varepsilon}{\beta^2} \ ,
  \end{equation}
  in terms of which, the (\ref{del}) becomes
  \begin{eqnarray}\label{dlphi}
  \Delta\phi&=&\frac{2}{\alpha}\int_{0}^{1} \frac{dt}{\sqrt{(1-t^2)(1-mt^2)}}\nonumber \\ 
  \Delta\phi&=& \frac{2K(m)}{\alpha} \ .
  \end{eqnarray}
 The above expression is valid both for the WH case ($\epsilon<0$) and the GM case ($\epsilon>0$). Remembering that $\beta$ is the turning point, we will have $0<m<1$ for the WH; in this case $K(m)$ is an incomplete elliptic integral of the first type, given in terms of the parameter itself rather than the modulus. In the GM case, we can write $m=-\frac{\varepsilon}{\beta^2}=-a$, with $a=\frac{\varepsilon}{\beta^2}>0$. We can show, Appendix \ref{ap}, that
 \begin{equation}
 	K(m)=K(-a)=\frac{1}{\sqrt{a+1}}K\bigg(\frac{a}{a+1}\bigg) \ .
 \end{equation}
    
\begin{figure}[h]
  	\centering
  	\includegraphics[width=\columnwidth]{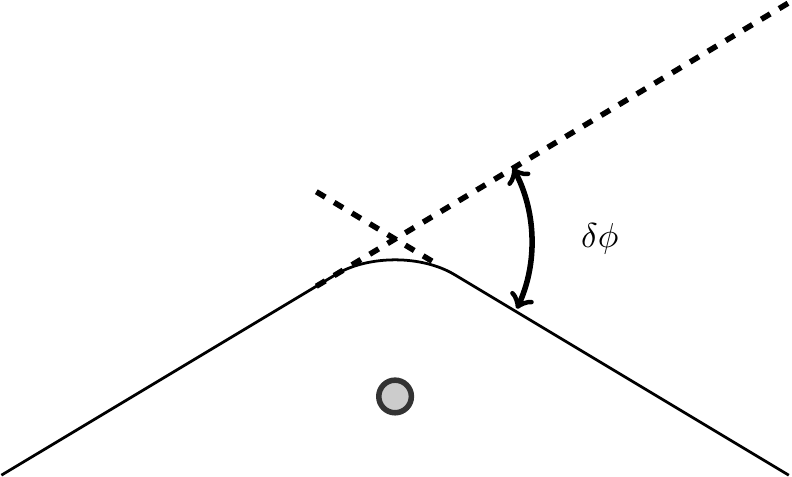}
  	\caption{Scheme of the deflection of light, $\delta\phi$, due to gravitational attraction.} 
  	\label{LG}
  \end{figure}

  The light deflection, angle between the new direction of propagation and the previous direction, as shown in the Fig. \ref{LG}, it is given by $\delta\phi=\Delta\phi-\pi$. Therefore,
  \begin{equation}\label{da}
  	\delta\phi=\frac{2K(m)}{\alpha}-\pi \ .
  \end{equation} 
 
   The expression (\ref{da}) is valid for both $m>0$ and $m<0$. But for didactic reasons we will discriminate, below, the deflection for a WH and a GM.
   
    $$\delta\phi=\begin{cases}
    \frac{2K(-\varepsilon/\beta^2)}{\alpha}-\pi,\quad\mbox{($\varepsilon<0$, WH)},\\
    
    \frac{2}{\alpha\sqrt{(\varepsilon/\beta^2)+1}}K\bigg(\frac{(\varepsilon/\beta^2)}{(\varepsilon/\beta^2)+1}\bigg)-\pi,\quad\mbox{($\varepsilon>0$, GM)} \ .
    \end{cases}$$\label{e}
    
\begin{figure}[h]
   	\centering
   	\includegraphics[width=\columnwidth]{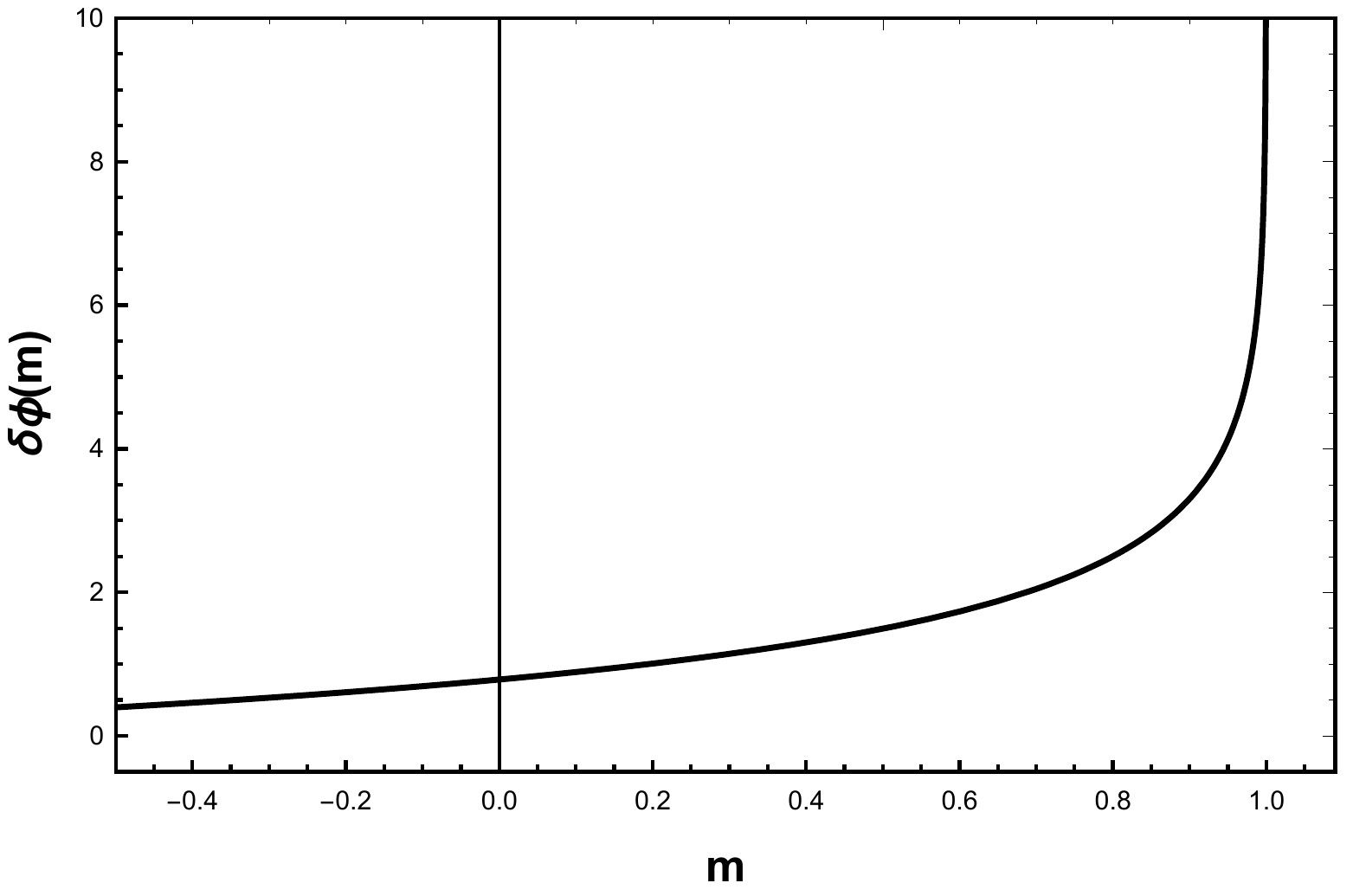}
   	\caption{Light deflection, $\delta\phi$, as a function of $m$, for $\alpha=0.8.$} 
   	\label{k}
   \end{figure}

  Of course, if $\varepsilon=0$ and $\alpha=1$, we will have $\delta\phi=0$, which corresponds to the case of flat spacetime, where there is no angular deflection since the gravitational attraction is null. In fig.\ref{k}, we plot the deflection for $\alpha=0.8$ in order to clarify some aspects related to lensing in the GM/WH spacetime. Note that in the WH case ($m>0$), when the turning point tends to the throat radius the deflection diverges, we call this limit  strong field. On the other hand, in the case of the modified GM ($m<0$), the deflection is always finite.

  First, let us investigate the deflection in the weak field limit, i.e., for $\beta\gg\varepsilon$, where $m\to0$. In this limit,  Eq.(\ref{da}) can be written as
  \begin{equation}\label{fraco}
  	\delta\phi=\frac{\pi(1-\alpha)}{\alpha}-\frac{\pi}{4\alpha}\Big(\frac{\varepsilon}{\beta^2}\Big)+\mathcal{O}\big(\frac{\varepsilon}{\beta^2}\big)^2 \ .
  \end{equation}
 The first term in the expression above corresponds to the deflection of a pure GM \cite{Barriola-Vilenkin}, and the second term is a contribution from EiBI gravity. We emphasize that in the WH case ($\varepsilon<0$), the deflection clearly becomes more pronounced than in the Barriola and Vilenkin GM case, the opposite occurs for EiBI GM. For the EiBI GM case, the deflection was first obtained by \cite{lambaga2018}. \\
  
  \section{Wormhole Light deflection in the strong field limit}\label{sec3}
  
\begin{figure}[h]
     	\centering
     	\includegraphics[width=\columnwidth]{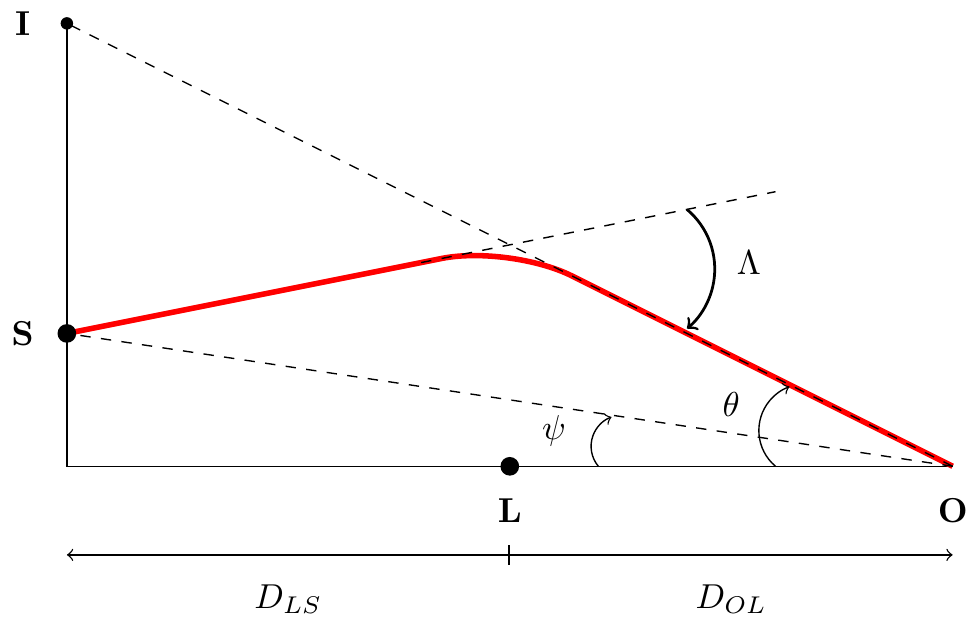}
     	\caption{Lens diagram.} 
     	\label{de}
     \end{figure}

   As already discussed,  Eq.(\ref{fraco}) is valid for both the WH and the GM. However, as we can see in Fig. \ref{k}, the light deflection never diverges in the GM case ($\varepsilon>0$). On the other hand, the deflection diverges for the WH case ($\varepsilon<0$) when $m\to1$. This limit, which corresponds to the approximation of the turning point towards the wormhole throat, is called the strong field. It is on this limit that we will focus from now on. For simplicity, let's take $\varepsilon=-a^2$; therefore, the parameter $a$ corresponds to the wormhole throat. Thus,  from (\ref{da}), we are left with
   \begin{equation}\label{forte}
   		\delta\phi=\frac{2}{\alpha}K\bigg(\frac{a^2}{\beta^2}\bigg)-\pi \ .
   \end{equation}
   In the strong field limit, $\beta\to a$, consequently, $\frac{a^2}{\beta^2}\to 1$. According to \cite{Soares2020}, in this limit, the deflection of light diverges logarithmically, that is \footnote{We could also get $\delta\phi$ in the strong field limit, taking into account that $
   	\lim\limits_{m\to1}K(m)=-\frac{1}{2}\log(1-m)+\frac{3}{2}\log(2)+\mathcal{O}{(1-m)\log(1-m)}$ \cite{booktra}.},
   \begin{eqnarray}\label{d}
   	\delta\phi&=&-\frac{1}{\alpha}\log\bigg(\frac{\beta}{a}-1\bigg) + \frac{3}{\alpha}\log(2)-\pi\nonumber\\
   	&&+\mathcal{O}{(1-a/\beta)\log(1-a/\beta)} \ .
   \end{eqnarray}
   With this result, we can study gravitational lensing in the strong field limit. To that end, let us review the lens equations at this limit in order to clarify our findings. In Fig.\ref{de} we present the visual profile of lensing. The light emitted by a source ({\bf S}) is deflected by the lens ({\bf L}), the wormhole, towards the observer ({\bf O}). With respect to the optical axis ({\bf LO}),  $\psi$ and $\theta$ give the angular position of the source and the image ({\bf I}), respectively. Evaluating the figure, it can be shown that \cite{Virbhadra-Ellis-2000}:
   \begin{equation}
   	\tan\psi=\tan \theta-\frac{D_{LS}}{D_{OS}}[\tan\theta-\tan(\Lambda-\theta)] \ ,
   \end{equation}
   where $D_{OS}=D_{OL}+D_{LS}$.
     
    Let's assume that the source and lens are almost perfectly aligned. In this sense, although the angular positions of the source and the image are small, the light ray circles the source several times before heading towards the observer, thus, $\Lambda$ must be very close to a multiple of $2\pi$ \cite{Boz-Cap2001}. Therefore, we can write $\Lambda=2\pi n+\Delta\Lambda_n$, where $\Delta\Lambda_n$ is the deflection angle after the $n$ loops around the lens, so $\tan(\Lambda-\theta)\sim\Delta\Lambda_n-\theta$. With these observations, the lens equation becomes
    \begin{equation}\label{lens equation}
    	\psi=\theta-\frac{D_{LS}}{D_{OS}}\Delta\Lambda_n \ .
    \end{equation}
   Still, in agreement with what we have stated, we can then write the critical impact parameter as
   \begin{equation}\label{ang}
   	\beta=D_{OL}\theta \ .
   \end{equation}
   With this, we can rewrite the angular deflection (\ref{d}) as
   \begin{equation}\label{def1}
   		\Lambda(\theta)=-\frac{1}{\alpha}\log\bigg(\frac{D_{OL}\theta }{a}-1\bigg) + \frac{3}{\alpha}\log(2)-\pi \ .
   \end{equation}
     We must remember, however, that whoever enters the lens equation (\ref{lens equation}) is $\Delta\Lambda_n$. To get $\Delta\Lambda_n$, we expand $\Lambda(\theta)$ em $\theta=\theta^{0}_n$, where $\Lambda(\theta^{0}_n)=2\pi n$, i.e.,
     \begin{eqnarray}\label{pa}
     	\Lambda(\theta)&=&\Lambda(\theta^{0}_n)+\frac{\partial\Lambda}{\partial\theta}\Bigg|_{\theta=\theta^{0}_n}(\theta-\theta^{0}_n)\nonumber\\
     \Delta\Lambda_n&=&\frac{\partial\Lambda}{\partial\theta}\Bigg|_{\theta=\theta^{0}_n}(\theta-\theta^{0}_n) \ .
     \end{eqnarray}
    Taking $\Lambda(\theta^{0}_n)=2\pi n$ in (\ref{def1}), we find
    \begin{equation}\label{t0}
    	\theta_{n}^{0}=\frac{a}{D_{OL}}\left[1+\frac{8}{e^{(1+2n)\alpha\pi}}\right] \ .
    \end{equation}
    Substituting (\ref{def1}) and (\ref{t0}) in (\ref{pa}), we are left with 
    \begin{equation}\label{dln}
    	\Delta\Lambda_n=-\frac{e^{(1+2n)\alpha\pi}D_{OL}}{8\alpha a}\Delta\theta_n \ .
    \end{equation}
   Substituting (\ref{dln}) in the lens equation (\ref{lens equation}), we are left with
    \begin{equation}\label{ir}
    	\theta_n\simeq\theta^0_{n}+\frac{8\alpha a D_{OS}(\psi-\theta^0_n)}{D_{LS}D_{OL}e^{(1+2n)\alpha\pi}} \ .
    \end{equation}
     The Eq.(\ref{ir}) gives the angular position of the $n$th relativistic image. See that they are influenced by the topological charge and when $\alpha\to1$, they fall back on the case of Ellis-Bronnikov \cite{Tsukamoto2016}. In general, $\alpha<1$, which implies that in the case of the topologically charged, the respective angular positions are greater than in the case of the Ellis-Bronnikov WH, which has no topological charge \cite{Tsukamoto2016}.
    
    The total flux from the $n$th lensed image is proportional to the magnification $\mu_n$, which is given by $\mu_n=\Big[  \frac{\psi}{\theta}\frac{\partial\psi}{\partial\theta}\bigg|_{\theta_{n}^{0}}\Big]^{-1}$ \cite{Boz-Cap2001}. Therefore, from (\ref{ir}) and (\ref{t0}),we can show that
   \begin{eqnarray} \label{mag}
    	\mu_n&=&\frac{8\alpha D_{OS}}{\psi D_{LS}} \Big(\frac{a}{D_{OL}}\Big)^2 \frac{1}{e^{(1+2n)\alpha\pi}} \nonumber\\
    	&&\times\left[1+\frac{8}{e^{(1+2n)\alpha\pi}}\right] .
   \end{eqnarray}
   
    As we can see, the magnification decreases rapidly with $n$, indicating that the brightness of the first image, $\theta_1$, is higher than the others. On the other hand, since in general $D_{OL}\gg a$, the magnification is always small. We can also observe that the greater the alignment between the source and the lens ($\psi \ll1$), the stronger the magnification. 
    
    The equations (\ref{ir}) and (\ref{mag}) express the positions of the relativistic images and the magnifications in terms of the parameters that characterize the wormhole ($a$ and $\alpha$). We can think in reverse, that is, we can define observables that can be measured via observational methods. Since these observables can be resolved to obtain the parameters that characterize the wormhole, they can present themselves as a good way to search for wormholes. In addition, the such methodology can also be useful as a tool for research beyond GR, since from the observables we can discriminate between a GR WH and an EiBI WH. In \cite{Bozza2002}, Bozza defined the following observables:
   \begin{equation}
   	s=\theta_1-\theta_{\infty}\quad\text{and}\quad R=\frac{\mu_1}{\sum_{n=2}^{\infty}\mu_n} \ .
   \end{equation}
   Where $s$ is the angular separation between the first and the rest of the relativistic images and  $R$   is the relationship between the flux of the first image and the flux of all the others.
    From (\ref{ir}), we can show that $s$ which is given by 
    \begin{equation}\label{ss}
    	s\simeq\frac{8 \theta_\infty}{e^{3\alpha\pi}} \ ,
    \end{equation}
    and from (\ref{mag}), we can also show that
    \begin{equation}\label{R}
    	R=\frac{\left(e^{-3\pi\alpha}+8e^{-6\pi\alpha}\right)}{e^{-5\pi\alpha}+8e^{-10\pi\alpha}} \ .
    \end{equation}
    For $\alpha\simeq 1$, as we believe it should be, the expression (\ref{R}) becomes
    \begin{equation}\label{R1}
    	R\simeq e^{2\pi\alpha} \ .
    \end{equation}
    Compared to the Ellis-Bronnikov WH, the presence of the topological charge $\alpha$ decreases the value of the observable $R$ and increases the observable $s$. Please, compare equations (\ref{R1}) and (\ref{ss}) with equations (5.15) and (5.14) of \cite{Tsukamoto2016}, respectively.
    
    \section{Wormhole Lensing observational Aspects}\label{sec4}
    In order to have an observational perspective of the EiBI WH, we will model three astrophysical scenarios, one for the strong field limit and two for the weak field limit; the choices were made in order to make the data clearer and avoid too many approximations. First, let us assign well-motivated values to the parameters $a$ and $\alpha$. Recent works \cite{jana}, involving the velocity of propagation of gravitational waves, estimated an upper limit for the Eddington parameter ($\varepsilon$) that is, $|\varepsilon|\leq10^{37}m^2$. Remembering that the radius of the wormhole throat treated here is $a=\sqrt{|\varepsilon|}$, it is estimated that $a\leq10^{15} km$. Furthermore, based on grand unification theories \cite{Barriola-Vilenkin,Soares2019}, we can estimate $\alpha\sim\sqrt{1-10^{-5}}$.\footnote{According to section \ref{sec2}, $\alpha=\sqrt{1-\kappa^2\eta^2}$. On the other hand, as argued in \cite{Barriola-Vilenkin}, for a typical grand unification scale it is estimated that $\eta\sim10^{16}$Gev, which demonstrates the reasonableness of our estimate.}
    \subsection{Strong field limit}
    At the strong field limit, we modeled the wormhole with data from Sagittarius A*, at the center of our galaxy. The mass of SgrA* is estimated at $4.4\times10^{6}M_{\odot}$ and is at an approximate distance of $D_{OL}=8.5$Kpc \cite{Gezel2010}.  As we have discussed before, it is estimated that the radius of the EiBI WH throat has an upper bound, $a\leq10^{15}$ km. In table \ref{t0} we present the values of the observables, $\theta_{\infty}$, $s$ and $\tilde{R}$, for some values of the throat radius $a$. The observable $\theta_{\infty}$ is defined as the critical angle, of (\ref{ang}), $\theta_{\infty}=\frac{\beta}{D_{OL}}$. The observable $s$ is given by (\ref{ss}) and $\tilde{R}=2.5\log_{10}R$, where $R$ is given by (\ref{R1}), the latter redefinition is useful to compare our results with those found in the literature.
    \begin{table}[]
    	\caption{ Observable}
    	\label{t0}
    	\begin{tabular}{|l|l|l|}
    		\hline
    		$a(km)$ & $\theta_{\infty}$($\mu$ arcsecs)             & s($\mu$ arcsecs) \\ \hline
    		$10^9$        & $0.78634$    & $0.000507$              \\ \hline
    		$10^{10} $   & $7.8634$  &  $0.00507$               \\ \hline
    		$10^{11}$    & $78.634$   & $0.0507$                \\ \hline
    		$10^{12}$ & $786.34$   & $0.507$  \\ \hline
    	\end{tabular}
    \end{table}
   In the case of the well-known Schwarzschild black hole, the values of these observables for the scenario considered here are already known and are given by \cite{Virbhadra-Ellis-2000}: $\theta_{\infty}$=26.547$\mu$arcsecs, $s=0.03322$$\mu$arcsecs e $\tilde{R}=6.821$ magnitudes. According to table \ref{t0}, for $10^{10} km\leq a \leq 10^{11} km$, the order of magnitude of the critical angle and angular separation have the same order of magnitude as in the Schwarzschild case, but differ significantly for other values. In fact, the data indicate that for $a\ge10^{11}$ the values of the observables assume values more compatible with the current observational range than in the Schwarzschild case. As for the magnification (\ref{R1}), $\tilde{R}$, we observe that it does not depend on the throat radius, and is given by $\tilde{R}=6.821$ magnitudes; we note that it is similar to the Schwarzschild case. However, we must emphasize that using the critical angle and angular separation we can distinguish between the space of a Schwarzschild black hole and an EiBI WH using the strong field limit lensing.
   
    \subsection{Weak field limit}
    In the weak field limit, $a\ll1$ and $a/\beta \ll1$, there are no loops and the deflection is given by (\ref{fraco}), which for feasible topological charge values, becomes
    \begin{equation}\label{lambe}
    	\Lambda \sim  \frac{\pi(1-\alpha)}{\alpha}+\frac{\pi}{4\alpha}\Big(\frac{a^2}{\beta^2}\Big) \ .
    \end{equation}
    It is worth remembering that, as we are dealing with the wormhole case, we are taking $\epsilon=-a^2$ in (\ref{fraco}).
    In perfect alignment, $\psi=0$. The angular position, $\theta_E$, can be obtained from (\ref{lens equation}), (\ref{ang}), and (\ref{lambe}) which leads to the equation 
    \begin{equation}
    	\theta_E^3+\frac{D_{LS}\pi(1-\alpha)}{D_{OS}\alpha}\theta^2-\frac{D_{LS}\pi a^2}{D_{OS}4\alpha D_{OL}^2}=0 \ ,
    \end{equation}
    whose real solution for values of $\alpha$ close to unity is given by
    \begin{eqnarray}
    	\theta_E&=&\Bigg[\frac{\pi D_{LS}a^2}{4\alpha D_{OS} D_{OL}^2}\Bigg]^{1/3}\nonumber\\ &+&\frac{(1-\alpha)}{6D_{OS}}\Bigg[\bigg(\frac{a^2D_{LS}D_{OS}^22\pi}{D_{OL}^2}\bigg)^{1/3}-2\pi D_{LS}\Bigg].
    \end{eqnarray}
    From (\ref{ang}), we can calculate the Einstein radius, $R_E$,
    \begin{equation}
    	R_E=D_{OL}\theta_{E} \ .
    \end{equation}
      Let us now estimate the observables $R_E$ and  $\theta_E$ taking into account reasonable values for the model parameters. Following the example of \cite{Abe2010}, let's consider the lensing of  a bulge star and a  star in the Large Magellanic Cloud (LMC). For a bulge star, the following parameters are adopted: $D_{OS}=8kpc$ and $D_{OL}=4kpc$; while for the Large Magellanic Cloud: $D_{OS}=50kpc$ and $D_{OL}=25kpc$.    Based on these estimates, in the tables \ref{t1} and \ref{t2} we present some values of the observables for the two presented scenarios. We verified that, within the possibilities of EiBI gravity, the present wormhole presents feasible theoretical possibilities of being detected for values of the throat radius of the order of $10^9 km$ or more, which is well accommodated within the restrictions for the parameters of the mentioned gravity theory. The angular deflection and the angle of the Einstein ring is of the order of 10 arcsec, for realistic parameter values, and are certainly within the observable range.
    
\begin{table}[]
	\caption{Einstein Radii/angle for Bulge  Lensings}
	\label{t1}
	\begin{tabular}{|l|l|l|}
		\hline
		$a(km)$ & $R_E(km)$             & $\theta_E(mas)$ \\ \hline
	  $10^9$        & $4.14\times10^{10}$    & 69.31               \\ \hline
		$10^{10} $   & $1.36\times 10^{12}$  &  $2.28\times10^{3}$               \\ \hline
		$10^{11}$    & $7.53\times 10^{12}$   & $1.25\times10^{4}$                \\ \hline
		$10^{12}$ & $3.61\times 10^{13}$   & $ 6.04\times 10^{4} $  \\ \hline
	\end{tabular}
\end{table}

\begin{table}[]
	\caption{Einstein Radii/angle for LMC   Lensings}
	\label{t2}
	\begin{tabular}{|l|l|l|}
		\hline
		$a(km)$ & $R_E(km)$             & $\theta_E(mas)$ \\ \hline
		$6\times10^9	$       & $1.98\times10^{11}$    & 53.95                \\ \hline
		$10^{10} $   & $1.09\times 10^{12}$  &  $2.93\times 10^{2}$               \\ \hline
		$10^{11}$    & $1.24\times 10^{13}$   & $3.32\times 10^{3}$               \\ \hline
		$10^{12}$ & $6.51\times 10^{13}$   & $ 17.41\times 10^{3} $  \\ \hline
	\end{tabular}
\end{table}

\section{Conclusions and Perspectives}\label{conc}

In this work, we explore the  gravitational lensing of a GM/WH spacetime within EiBI gravity. Initially, we calculate the light deflection in a general and exact way, both for the case of a GM and for the case of a WH, and we emphasize that the presence of a topological charge allows observationally to discriminate whether such a solution is described by the RG or by the EiBI gravity. We emphasize that in the latter case, the solution is physically more attractive as it does not presuppose any source of exotic matter \cite{Soares2020}. For the time being, we conjecture that a topologically charged wormhole may arise within EiBI gravity as a result of the evaporation of a topologically charged black hole, we hope to publish our theoretical research in this direction soon. In the  weak field limit, we show that the light deflection increases for a wormhole and decreases for a GM compared to the Barriola and Vilenkin GM. We argue that in the case of GM the light deflection never diverges, on the other hand, in the case of WH, the light deflection logarithmically diverges in the strong field limit, that is, when the impact parameter tends to the radius of the WH throat. We studied the lensing of the EiBI WH in the strong field limit and showed that this methodology allowed us to distinguish between an EiBI WH and an Ellis-Bronnikov WH. We also performed an analysis of the observables taking into account plausible and well-motivated values for the parameters. The data suggest that at the weak field limit the observables are within the observational range.
 Finally, we also emphasize that the solution studied in this paper can collaborate in the understanding of optical properties of liquid crystals, since the spatial section of (\ref{eq2}) adequately describes these condensed matter systems \cite{Moraes103,Fumeron}.

 \appendix
\section{ $K(m)$ Function for  $m$ negative values.}\label{ap}
From (\ref{dlphi}), we know that 
\begin{equation}
	K(m)=\int_{0}^{1} \frac{dt}{\sqrt{(1-t^2)(1-mt^2)}} \ .
\end{equation}
Making the following substitution: $t=\sqrt{1-u^2}$, we are left with
\begin{equation}
	K(m)=\int_{0}^{1} \frac{du}{\sqrt{(1-u^2)(1-m+mu^2)}} \ .
\end{equation}
Therefore,
\begin{eqnarray}
K(-m)&=&\int_{0}^{1} \frac{du}{\sqrt{(1-u^2)(1+m-mu^2)}} \nonumber\\
K(-m)&=&\frac{1}{\sqrt{m+1}}\nonumber\\
&&\times
\int_{0}^{1} \frac{du}{\sqrt{(1-u^2)(1-\frac{m}{m+1}u^2)}} \ .
\end{eqnarray}
Then,
\begin{equation}
	K(-m)=\frac{1}{\sqrt{m+1}}K\bigg(\frac{m}{m+1}\bigg) \ . \ \square
\end{equation}


\end{document}